\documentclass[AMA,LATO1COL]{WileyNJD-v2}

\articletype{}

\received{}
\revised{}
\accepted{}

\graphicspath{ {./Images/} }

\usepackage{url}

\begin{document}

\title{Designing False Data Injection Attacks Penetrating AC-based Bad Data Detection System and FDI Dataset Generation}

\author{Nam N. Tran}
\author{Hemanshu R. Pota}
\author{Quang N. Tran}
\author{Xuefei Yin}
\author{Jiankun Hu*}

\authormark{Nam N. Tran \textsc{et al}}

\address{\orgdiv{School of Engineering and Information Technology}, \orgname{University of New South Wales Canberra at ADFA}, \orgaddress{\state{Australian Capital Territory 2620}, \country{Australia}}}

\corres{*Jiankun Hu, School of Engineering and Information Technology, University of New South Wales Canberra at ADFA, Australian Capital Territory 2602, Australia. \email{j.hu@adfa.edu.au}}


\abstract[Summary]{The evolution of the traditional power system towards the modern smart grid has posed many new cybersecurity challenges to this critical infrastructure. One of the most dangerous cybersecurity threats is the False Data Injection (FDI) attack, especially when it is capable of completely bypassing the widely deployed Bad Data Detector of State Estimation and interrupting the normal operation of the power system. Most of the simulated FDI attacks are designed using simplified linearized DC model while most industry standard State Estimation systems are based on the nonlinear AC model. In this paper, a comprehensive FDI attack scheme is presented based on the nonlinear AC model. A case study of the nine-bus Western System Coordinated Council (WSCC)'s power system is provided, using an industry standard package to assess the outcomes of the proposed design scheme. A public FDI dataset is generated as a test set for the community to develop and evaluate new detection algorithms, which are lacking in the field. The FDI's stealthy quality of the dataset is assessed and proven through a preliminary analysis based on both physical power law and statistical analysis.}


\maketitle

\section{Introduction}
\label{introduction}
The prominence of deploying information and communication technology in the traditional power system integrated with Distributed Energy Resources (DERs) such as solar, wind, energy storage, etc., has gradually revolutionized this infrastructure into a much more advanced cyber-physical system, called smart grid. Such transformation offers various benefits to the task of system monitoring and management which is undertaken by the Energy Management Systems (EMSs). It also introduces many new kinds of cyber-threats to the system. The core part of an EMS is the State Estimation (SE), which is assigned to do multiple tasks, including collecting and processing measurements, removing inadequate data, and solving an optimization problem to yield a set of state variables that closely reflect the status of the system. As the whole process involves data transmission over communication networks, problems of measurement data being compromised are emerging and getting more and more attention in the field.

Among all kinds of threats, a new type called False Data Injection (FDI) attack is considered extremely dangerous as it is completely stealthy under examination by the State Estimation, as shown in Liu et al. \cite{liu2011false} and Hug et al.\cite{hug2012vulnerability}. FDI attacks aim directly at the set of measurements by injecting manipulated data, thus driving the system's behavior in a malevolent way. From the view point of the adversaries, this man-in-the-middle type of attack is an effective method to mislead the operators at the control center. For instance, an FDI attack might lead the SE to indicate some surges of power flow on lines, which could make the operators to think about the malfunctioning of the automatic relay protection system. As a result, the operators might establish an emergency protective mechanism to trip down the suspected lines, which might possibly bring out power outages. Since, the discontinuity of power supply is a severe incident due to its consequences, it is apparent that FDI attack can inflict damage to the physical system as well. In another situation, an FDI attack might help hackers to take advantage by changing power consumption at some specific nodes, which will result in significant errors in load forecasting and generation planning.

Most of existing research works in this field are based on the simplified DC model, e.g., Liu et al. \cite{liu2011false}, Bobba et al. \cite{bobba2010detecting}, and Anwar et al.\cite{anwar2017modeling}. However, it is important for researchers and industrial professionals to understand the FDI's stealthy capability under the AC-based practical setting model, which has been widely deployed in the industry. In this paper, a comprehensive FDI attack design scheme is presented to examine its effectiveness using fully nonlinear AC power flow model. Thus, it is much more challenging than most of the previous works which were purely based on simplified linearized DC-based models. The proposed FDI attack scheme has been tested on a case study to assess its capability to penetrate a commercial-level AC based SE package, PowerFactory 2017 SP4 from DIgSILENT\cite{PF2017}. As demonstrated by the experimental results, our proposed method of FDI attack has successfully exploited a critical vulnerability in the state-of-the-art Bad Data Detector (BDD) and plausibility checking. As both the existing detection methods with electrical (for plausibility checking) and stochastic (for bad data removing) basis have failed to accomplish their missions, new FDI attack detection mechanisms are needed. Unfortunately there exist no realistic AC-based FDI attack dataset publicly available, which has greatly hindered the relevant technology development in terms of designing new FDI attack detection mechanisms and their evaluation. In order to provide a practical tool to evaluate FDI detecting techniques, a cyber-physical dataset is constructed and released publicly. Unlike the case of pure cyberspace environment which has several datasets for testing design theories (for instance KDD \cite{KDD}, DARPA \cite{DARPA}, ADFA-LD \cite{asathbidsucadscp_creech, goanitdttrtkc_creech}, NGIDS-DS \cite{NGIDS-DS}, etc.), to the best of our knowledge, our FDI dataset is the first ever AC-based FDI dataset. It has been created with highly realistic settings as a result of a combination of the use of an industry-standard commercial-level package (PowerFactory) as the test platform, a trusted source of input data (Australian Energy Market Operator), and a sophisticated attack design process presented in this paper. We also provide some preliminary analyses in order to demonstrate the cause for the failure of BDD equipped with this SE package.

The primary contributions from this paper can be summarized as follows:
\begin{enumerate}
	\item Providing an evaluation, using a commercial SE package, of the performance of DC-based FDI attack model against an AC-based system. Note that most of the FDI detection schemes are based on DC-based FDI attack model and have been evaluated in a simplified simulation setting. It is important and also interesting to observe the behavior of the system in commercial-scale settings that are used in real systems.
	\item Proposing a systematic AC-based FDI attack design scheme. An experimental case study is provided, which shows the success of perfectly bypassing a BDD in commercial-level SE package.
	\item Generating the first cyber-physical FDI attack dataset for public research purpose.
	\item Providing preliminary analyses to explain the failure of the state-of-the-art BDD against our proposed attack scheme.
\end{enumerate}

The organization of this paper includes six sections with the current section being the introduction of the research problem. Section \ref{prevwork} reviews various related work and also provides an assessment of some popular DC-based attack schemes against AC-based SE. Section \ref{design} presents a comprehensive attack design scheme based on the AC model of power flows. Based on the proposed design scheme, a large-scale cyber-physical FDI attack dataset is generated which is described in Section \ref{dagen}. Section \ref{preana} provides preliminary analysis from both electrical and stochastic perspectives in order to highlight the stealth quality of the generated FDI dataset. Conclusion and some future researches are given in Section \ref{conclu}.

\section{Previous work}\label{prevwork}
Due to the advent of the pioneering work by Liu et al. \cite{liu2011false}, there have been myriads of publications on the FDI attack against SE in power system. Liu et al. \cite{liu2011false} comprehensively categorized almost every type of FDI attack. One limitation of this work is that the linearized DC-based power flow model is deployed to test the attacks. Despite the authors' claim to expand their future research into the field of nonlinear AC-based  model, most of, if not all, following works still applied the DC-based model. According to the review in Liang et al.\cite{liang2017review}, it pointed out that in the field of FDI attack research including types, impacts and defense strategies, DC-based publications still dominate the literature. 

Sharing the same foundation of DC power flow model, Yang et al.\cite{yang2014false} proposed an optimized way to attack with minimum effort, and Rahman et al.\cite{rahman2012false} confirmed that an attack without complete knowledge about the system is possible. Similar to the work of Yu et al.\cite{yu2015blind} where a proposed attack scheme was based on the Principal Component Analysis to aim at creating a blind FDI attack, Esmalifalak et al. \cite{esmalifalak2011stealth} assumed loads are invariant and then formed an attack based on the Independent Component Analysis for stealthy purpose. Liu et al. \cite{liu2015modeling} introduced a strategy to determine optimal attack region given the limited network information. Meanwhile, Adnan et al. \cite{anwar2017modeling} tried to construct an attack by building up low-rank original measurement matrix from observed measurement matrix. Kim et al. \cite{6810522} introduced a variant of FDI attack named Data Framing which is unable to completely bypass the BDD but is sophisticated enough to deceive the Bad Data Identification (BDI), thus mislead the SE to produce incorrect system's states. During this period, several research groups also published various defense strategies against FDI attack, including heuristic \cite{kosut2010limiting}, protecting right from the measurement \cite{bobba2010detecting}, Sparse Optimization \cite{liu2014detecting}, using sequential detector \cite{li2015quickest} or adaptive CUSUM test \cite{huang2011defending} and even a graphical method \cite{bi2014graphical}. All of these defense schemes are planned with great details and some even mention the financial consequences \cite{xie2010false}. These results have greatly contributed to the advancement in this field. However, whether or not such works are applicable for the AC-model based realistic systems is an open question.

For its simplicity, there is no doubt that the DC-based model is a good starting point for FDI research. With the development of computing technology, the problem of computational cost or convergence issue can no longer hinder the implementation of AC-based SE in the EMS. By removing several simplified assumptions, AC-based model can provide a better estimated set of results. This fact can be easily confirmed once we investigate the fundamental power flow equations and measurement model for each type:
\begin{itemize}
	\item DC power flow equation: $P_{ij} = \frac{\theta _i - \theta _j}{X_{ij}}$.
	\item DC measurement model: $z = \mathbf{H}x + \epsilon$.
	\item AC power flow equation: $P_{ij} = f(V_i, V_j, g_{ij}, b_{ij}, \theta _i, \theta _j)$
	\item AC measurement model: $z = \mathbf{H}(x) + \epsilon$
\end{itemize}
The difference between these two models is quite significant as the AC model also takes into account the equation for reactive power flows. Consequently, it is highly possible that any designated technique for DC-based model will not work well with an AC-based SE. For illustration purpose, we conducted a test with the WSCC's 9-bus system following the primitive attack method against the SE package in PowerFactory 2017. As claimed by Adnan et al. \cite{anwar2014vulnerabilities}, a random contaminated vector \textit{c} in conjunction with the system Jacobian matrix \textbf{H} whose elements are partial derivatives of power flow equations, will be enough to produce an attack vector $a =$ \textbf{H}$c$ that could successfully bypass BDD in an SE. Assuming the \textbf{H} matrix has already been acquired, we randomly generated 100 different vectors \textit{c} to create 100 sets of corrupted measurements accordingly. After feeding measurement values respectively into the PowerFactory's SE, we achieved two types of outcomes: 46 cases failed to converge while 54 cases converged successfully but none has been able to completely bypass the BDD as expected. Moreover, the existence of manipulated measurements also resulted in the false alarm of some other good measurements. From these experimental results, designing FDI attack derived from DC power flow model is unlikely to possess its stealthy characteristic against an industry-standard AC-based SE. Such failures of DC-based design were also recognized in Hug et al.\cite{hug2012vulnerability}, Manandhar et al.\cite{manandhar2014detection}, Chaojun et al. \cite{chaojun2015detecting}, Liang et al.\cite{liang2016vulnerability} and Rahman et al. \cite{rahman2013false} as various methods were proposed to produce corrupted AC power flow model measurement values. As these works did not state explicitly  the examination environment to test their designs, the first target that has inspired our work is designing an attack to completely bypass the more realistic nonlinear AC model based BDD. In addition, it is equally important to generate the first FDI dataset as a benchmarking tool for conducting research in evaluating their proposed FDI detection solutions.

\section{Design scheme and Experimental result of an FDI Attack against AC State Estimation in Power System}\label{design}
\subsection{Design principle}
\begin{figure}[htbp]
	\centering
	\includegraphics[width=0.3\linewidth]{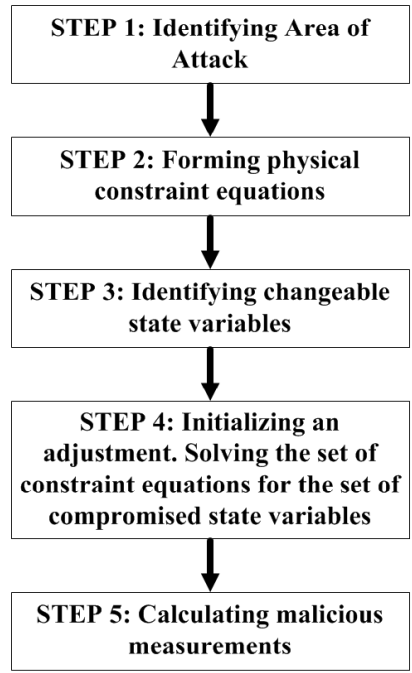}
	\caption{Details of Stage (\textit{s2}) in the FDI attack design scheme to generate an FDI attack - Design the malicious measurements.}
	\label{fig1}
\end{figure}

Technically, the transmission system for SE research is particularly considered as a quasi-static model whose variables are constantly yet slowly changing. The task of system monitoring is conducted in a discrete manner by the means of gathering data about the system, then building up a snapshot that reflects the current state. This process is repeated after an interval, generally about 5 to 10 minutes. From an attacker's point of view, the process to generate an FDI attack includes two main stages: (\textit{s1}) Collecting data about the current state of the system (or a subsystem of interest); (\textit{s2}) Designing the malicious measurement values and then injecting them into the data acquisition system. The second stage can be divided into five tasks, corresponding to: identifying Area of Attack, forming constraint equations, identifying changeable state variables, initializing adjustment and solving equations, and computing with the corrupted measurements. If the attackers already have enough information about the configuration of the system or subsystem, i.e. (\textit{s1}) is done, it will take almost no time to design a completely stealthy attack (\textit{s2}) with our proposed attack design scheme. In this paper, we will focus on the mission of systematically calculating the manipulated values of measurements (\textit{s2}, whose sub-stages are shown in Fig. \ref{fig1}) in order to completely bypass the existing BDD\cite{PF2017}.

Since this work mainly concentrates on the design process of malicious values, several assumptions are made in this study. Firstly, attackers can collect the data relevant to one state of the systems (or a subsystem of interest), including topology, status of the breakers and transformer taps, state variables (voltage magnitudes, voltage angles), power flows measurements (both active and reactive powers) on branches, and power injection measurements at nodes. Secondly, the attackers are able to not only retrieve data from the system but also overwrite the values sent by measurement devices to the control center. In addition, all the measurement devices are considered having the same accuracy and weight. Such a situation can arise when either the working computer or working account of the operator is compromised. Such assumptions have been used in most of FDI based applications \cite{hug2012vulnerability,manandhar2014detection,liang2016vulnerability}.

\subsubsection{Attack Area}
From the viewpoint of state estimation, the state of being injected by an FDI attack is equivalent to a new steady state, which consists of both genuine as well as manipulated measurement values. In other words, FDI attack will replace a set of system's measurements by another set that contains either only manipulated measurement values or both unchanged and corrupted values. In fact, it is reasonable for the attacker to only aim at adjusting a set of values in a subsystem or a sub-grid rather than taking risk of probably being detected due to attacking on a wide area. The smaller the affected area is, the higher the chance that the attack is hidden. The affected or attack area here is defined as a region that consists of all branches whose power flows are altered. Thus, the boundary of this region is formed by a set of nodes that are intersections of at least one affected branch to one or several branches with no corrupted power flows.

The first and foremost task in identifying the area of attack is finding its boundary. According to Hug et al.\cite{hug2012vulnerability}, the condition for a stealthy FDI attack is partly satisfied if the area of attack has the boundary of all power-injection nodes. Unlike the traditional way to categorize buses in load flow calculation, the FDI attack design process only separates bus type into two groups: (\textit{$B_1$}) bus \textit{with power injection} (either incoming or outgoing) and (\textit{$B_2$}) bus \textit{with no power injection}. The reason that the area of attack must be enclosed by only (\textit{$B_1$}) buses is due to the fact that an FDI attack alters the measurement values at both ends of every branch within the range of attack. Since, at every (\textit{$B_2$}) bus, the Kirchhoff's Current Law must be satisfied, this kind of bus will force at least one connected branch's power flows to be changed. Consequently, the existence of (\textit{$B_2$}) bus will result in the expansion of area of attack. Such expansion will be continued till there is no bus on the boundary belonging to (\textit{$B_2$}) type. For instance, consider two connected branches (\textbf{a-b}) and (\textbf{a-c}) with a common terminal \textbf{a} being of (\textit{$B_2$}) type. Any change of power flow happens on (\textbf{a-b}) will result in at least another couple of changes on branch (\textbf{a-c}) since the algebraic sum of power flows at node \textbf{a} must be equal to zero. In order to make these changes to be reasonable, the terminal \textbf{c} of the latter branch must have power injection there. An illustrative example about this statement is provided in Hug et al. \cite{hug2012vulnerability}.

In this work, the process of identifying FDI attack area is conducted by using an \textit{extended admittance matrix} $Y_{bus}$. The extended $Y_{bus}$ for an \textit{n}-bus system is a matrix that has size of $n\times(n+1)$. Its formation is based on the horizontal concatenation of the traditional admittance matrix $Y_{bus}$ with an additional column, which has the information about the type of the node. Elements of the additional column have values of either 0 or $\pm$1, depending upon whether a bus is of no-injection or power injection type. The bus that has power injection is further distinguished by a sign in order to support the process of automatic design attack, with +1 denoting a load and -1 denoting a bus with power going out. The Algorithm \ref{Algo1} presents the area of attack identification process.
\begin{algorithm}[htbp]
	\caption{Identifying Area of Attack}
	\label{Algo1}
	\begin{algorithmic}[1]
		\renewcommand{\algorithmicrequire}{\textbf{Input:}}
		\Require Extended $Y_{bus}$, Central node \textit{i}
		\renewcommand{\algorithmicensure}{\textbf{Output:}}
		\Ensure Area of Attack $\Omega_A$
		
		\State <\textbf{1}> Look for $i^{th}$ column of the extended $Y_{bus}$
		\State <\textbf{2}> Obtain $\Omega_i = \{j| Y_{ji} \neq 0\}$
		\State <\textbf{3}> Scanning process:
		\While{($j \in \Omega_i$)}
		\If{($Y_{j(n+1)} \neq 0$)}
		\If{(Last of $\Omega_i$)} \State $\rightarrow$ Jump to <\textbf{4}>
		\Else 
		\State \textit{j}++
		\EndIf
		\Else
		\State Do <\textbf{1}> - <\textbf{3}> with new central node \textit{j} (and its list $\Omega_j$)
		\EndIf 
		\EndWhile
		\State <\textbf{4}> Add up all $\Omega_i + \Omega_j ...$ to obtain the Area of Attack $\Omega_A$
	\end{algorithmic}
\end{algorithm} 

\subsubsection{Constraint Equations and Changeable State Variables}
The next step in the attack design process is setting up the set of physical constraint equations by the use of Algorithm \ref{Algo2}. Its solution is the essential resource for computing the manipulated measurements, which, in turn, will effectively bypass the BDD of AC-based SE. The number of the constraint equations depend on the configuration of the system or the subsystem, particularly, the number of no-injection nodes within the area of attack. The formation of the constraint equations originates from the Law of Conservation of Energy \cite{Wiki}, which offers a smooth transition from a steady state of operation (the genuine one) to another steady state of operation (the corrupted one), thus, guaranteeing the stealthy characteristic of the attack. Since the state of being attacked by FDI technique is essentially a counterfeit steady state, it must satisfy the following two constraints:
\begin{itemize}
	\item The sums of active and reactive power flows at any no-injection node must be equal to zero. Thus, for any no-injection Bus \textit{j}, we have: $\sum_{j} P_{jk} = 0$ \& $\sum_{j} Q_{jk} = 0$.
	\item All the changes in power injections (both active and reactive) at nodes as well as power losses (both active and reactive) on branches must add to zero. Therefore we have the constraint as: $\sum \Delta P_{INJ} + \sum \Delta P_{LOSS} = 0$ \& $\sum \Delta Q_{INJ} + \sum \Delta Q_{LOSS} = 0$.
\end{itemize}
\begin{algorithm}
	\caption{Forming physical constraint equations}
	\label{Algo2}
	\begin{algorithmic}[1]
		\renewcommand{\algorithmicrequire}{\textbf{Input:}}
		\Require Extended $Y_{bus}$, Central node \textit{i}, Area of Attack $\Omega_A$
		\renewcommand{\algorithmicensure}{\textbf{Output:}}
		\Ensure The set of constraint equations \textbf{S}
		
		\While{(($j \in \Omega_A$) and ($Y_{ij} \neq 0$))}
		\If{($Y_{j(n+1)} = 0$)}
		\State Write $\sum_j P_{jk} = 0$
		\State Write $\sum_j Q_{jk} = 0$
		\Else
		\State Write $\sum_j \Delta P_{INJ} + \sum_{jk} \Delta P_{LOSS} = 0$
		\State Write $\sum_j \Delta Q_{INJ} + \sum_{jk} \Delta Q_{LOSS} = 0$
		\EndIf
		\State \textit{j}++ 
		\EndWhile 
	\end{algorithmic}
\end{algorithm}

After constructing all the constraint equations, the next work is to identify which state variable will be changed, which is presented in the Algorithm \ref{Algo3}. The number of changeable state variables within an area of attack depends on the type of node, such as slack bus, PV bus, or load bus, within it. Therefore, if the area of attack with \textit{n} nodes contains a slack bus, which always has fixed value for both the voltage magnitude and voltage angle, it is apparent that the set of changeable state variables will be $(2\times n - 2)$. Likewise, for the existence of a PV bus in an area of attack with \textit{n} nodes, the set of changeable state variables will be $(2\times n - 1)$. Since the set of constraint equations is nonlinear, an iterative numerical technique and approximate solutions are expected. For that reason, the relationship between the number of equations and the number of unknown or changeable state variables is important. For the sake of simplicity in this work, we aim at a designing process with the number of constraint equations being equal to the number of unknowns. Thus, if the number of changeable state variables in the area of attack is greater than the number of constraint equations, not every value of state variable will be manipulated during the attack design process.
\begin{algorithm}
	\caption{Identifying changeable state variables}
	\label{Algo3}
	\begin{algorithmic}[1]
		\renewcommand{\algorithmicrequire}{\textbf{Input:}}
		\Require Area of Attack $\Omega_A$
		\renewcommand{\algorithmicensure}{\textbf{Output:}}
		\Ensure The set of changeable state variable \textbf{SV}
		
		\State No. of SV $n \leftarrow 2\times Size of(\Omega_A)$
		\State \textbf{SV} = \{$|V|_j, \theta _j| j \in \Omega_A$\}
		\While{($j \in \Omega_A$)}
		\If{(type of node (\textit{j}) = Slack)}
		\State \textit{n} = \textit{n} - 2
		\State Remove $|V|_j$ and $\theta _j$
		\ElsIf {(type of node (\textit{j}) = PV)}
		\State \textit{n} = \textit{n} - 1
		\State Remove $|V|_j$
		\Else
		\State \textit{j}++
		\EndIf
		\EndWhile 
	\end{algorithmic}
\end{algorithm}

As soon as the solution for the set of the constraint equations is obtained, all the malicious measurement values are ready to be computed using the traditional load flow formulas. Then, the complete set of measurements, or in other words the input data for the SE package, is formed by combining the set of valid measurements with the set of malicious measurements. The whole design process is illustrated with a case study of the WSCC's 9-bus system \cite{anderson2008power} in the next section.

\subsection{Case study - The WSCC's 9-bus Power System}
\begin{figure}[htbp]
	\centering
	\includegraphics[width=\linewidth]{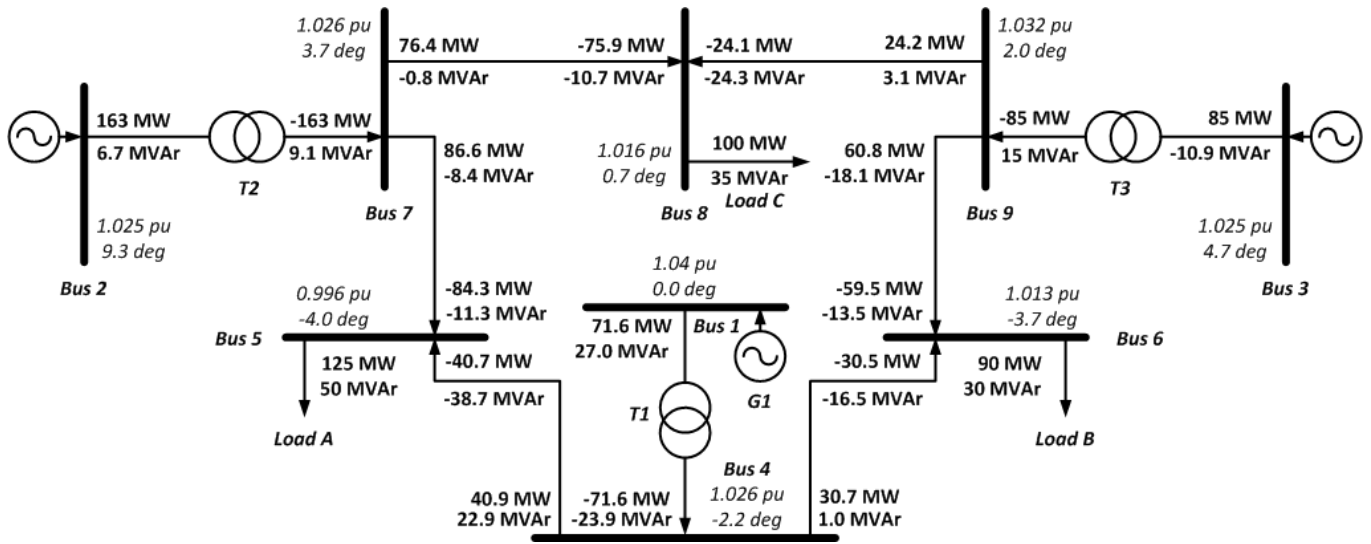}
	\caption{The nine-bus system and its load flow result.}
	\label{fig2}
\end{figure}

The WSCC's 9-bus power system \cite{anderson2008power} in Fig. \ref{fig2} is selected as the case study due to several reasons. With nine branches, three sources and three loads, it is large enough to represent a typical interconnected power system. More importantly, its datasheet provides us with adequate information about all generators. Some well-known IEEE benchmarking system such as IEEE 14-bus, IEEE 30-bus, IEEE 57-bus, etc., are all lacking of such data. In the current SE-oriented work, 48 measurements, with 24 active power measurements and 24 reactive power measurements, will be deployed. Among 24 measurements of each type, 18 are deployed to monitor power flows at each end of branches and 6 are used to acquire data of power injections at nodes. 

\begin{figure}[htbp]
	\centering
	\includegraphics[width=0.5\linewidth]{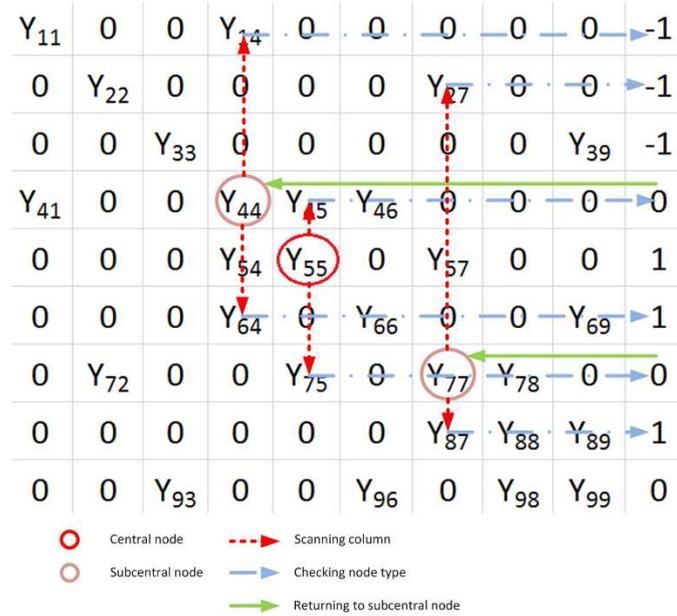}
	\caption{The extended $Y_{bus}$ matrix for the nine-bus system and the process of identifying area of attack with central node is Bus 5.}
	\label{fig3}
\end{figure}

The design scheme presented in the previous section is first applied here to illustrate the algorithm in finding the attack area. The extended $Y_{bus}$ for this nine-bus system is given in Fig. \ref{fig3}. For instance, an FDI attack is launched by choosing Bus 5 as the initial point. The detailed process is visually illustrated in Fig. \ref{fig3}, resulting in the area of attack including bus \{5, 4, 7, 1, 6, 8, 2\}. Assuming that this result is also the smallest possible area of attack, then we can move to the next phase of the design process.

For the nine-bus system, given the area of attack as identified above, the set of constraint equations will be formed as follows. Since there are two no-injection nodes within the area of attack, Bus 4 and Bus 7, the first constraint will render \textbf{four} equations in total, $\Sigma P_{4j} = 0$, $\Sigma Q_{4j} = 0$, $\Sigma P_{7j} = 0$, and $\Sigma Q_{7j} = 0$. In addition, the second constraint always renders \textbf{two} equations regardless of the configuration of the system, thus the set of constraints has \textbf{six} equations in total. The complexity of the two equations based on the second constraint depends upon the number of loads, sources and branches within the area of attack.  In this case, there are six branches, including both transformer branches and lines, as well as three loads and two sources. In total, there are eleven terms in each equation from the second constraint. The detailed formation of each equation from the set of constraints is presented as below.
\begin{enumerate}[1)]
\item The first constraint for the sums of active power and reactive power at no-injection Bus 4 and Bus 7
\begin{equation}
\label{eq1}
P_{41} + P_{45} + P_{46} = 0
\end{equation}
\begin{equation}
\label{eq2}
Q_{41} + Q_{45} + Q_{46} = 0
\end{equation}
\begin{equation}
\label{eq3}
P_{72} + P_{75} + P_{78} = 0
\end{equation}
\begin{equation}
\label{eq4}
Q_{72} + Q_{75} + Q_{78} = 0
\end{equation}
\item Parts of the second constraint for the sum of all changes in active  and reactive power injections within the attack area (Equation (\ref{eq5}) and Equation (\ref{eq6})) and for the sum of all changes in active and reactive power losses of all branches within the attack area (Equation (\ref{eq7}) and Equation(\ref{eq8}))
\begin{equation}
\label{eq5}
\Sigma \Delta P_{INJ} = \sum_{i=1, 2, 5, 6, 8} (P_i^{new} - P_i^{old})
\end{equation}
\begin{equation}
\label{eq6}
\Sigma \Delta Q_{INJ} = \sum_{i=1, 2, 5, 6, 8} (Q_i^{new} - Q_i^{old})
\end{equation}
\begin{equation}
\label{eq7}
\Sigma \Delta P_{LOSS} = \sum_{ij=14, 27, 45, 46, 57, 78} (P_{L,new}^{ij} - P_{L,old}^{ij})
\end{equation}
\begin{equation}
\label{eq8}
\Sigma \Delta Q_{LOSS} = \sum_{ij=14, 27, 45, 46, 57, 78} (Q_{L,new}^{ij} - Q_{L,old}^{ij})
\end{equation}
\end{enumerate}

In the current example of the nine-bus system, the area of attack with six equations has already been identified. Among all nodes within the area of attack \{1, 2, 4, 5, 6, 7, 8\}, nodes 6 and 8 are the nodes with power injection at the boundary. In order to limit the affected range to reach no further than the boundary, we want to keep their state variables unchanged (both the voltage magnitudes and the voltage angles). From the rest of the set \{1, 2, 4, 5, 7\}, node 1 is the slack bus and node 2 is a PV node. Consequently, there are \textbf{seven} adjustable state variables within the area of attack \{$\theta _2$, $\theta _4$, $\theta _5$, $\theta _7$, $V_4$, $V_5$, $V_7$\}. Meanwhile, the current principle of attack design is about first choosing one state variable as an initial point of attack then solving the set of constraint equations for manipulated state variables, which, in turn, will be used to calculate the malicious measurement values. Thus, the requirement for the number of changeable state variables is, in fact, one unit more than the number of the constraint equations. In the current example, Bus 5 is chosen as an initial point by arbitrarily adjusting its angle by an amount of 0.5 degree. The next task is to solve the set of six nonlinear equations \{Equation (\ref{eq1}), Equation (\ref{eq2}), Equation (\ref{eq3}), Equation (\ref{eq4}), (\ref{eq5}) + (\ref{eq7}) = 0, (\ref{eq6}) + (\ref{eq8}) = 0\} for the six unknowns \{$\theta _2$, $\theta _4$, $\theta _7$, $V_4$, $V_5$, $V_7$\}.

In order to evaluate the result of the above design scheme, SE package of PowerFactory is employed. This package is equipped with all critical features that an industry-standard State Estimation package requires, including but not limited to receiving input/measurement data from Remote Telemetry Units (RTUs), checking plausibility, solving an optimization problem to find estimated values of state variables, checking observability, marking redundant data as well as bad data, etc. For the purpose of testing our design theory, eight different cases of attack are prepared (corresponding to the change of Bus 5 voltage angle of $\pm$0.5, $\pm$1, $\pm$1.5, $\pm$2 degrees).

\begin{figure}[h]
	\centering
	\includegraphics[width=0.4\linewidth]{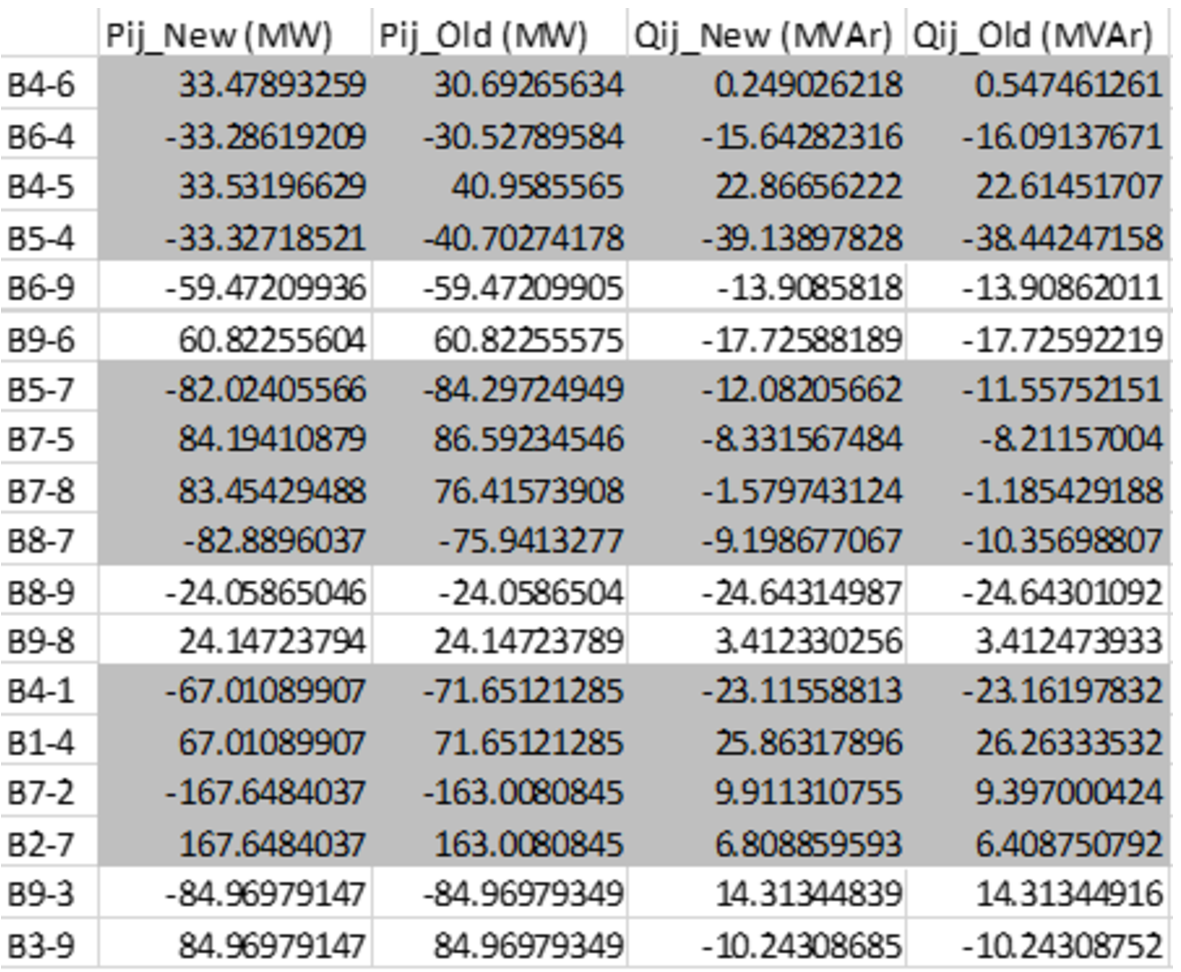}
	\caption{Comparison between the manipulated and the steady-state measurement values on branches.}
	\label{fig4}
\end{figure}

All the eight attack schemes successfully bypassed BDD of the SE module in the PowerFactory package without being detected even just one false measurement. For the case with an initial +0.5 degrees adjustment at Bus 5 voltage angle, 34 out of 48 measurements were intervened (marked by the cell with gray shade) in Fig. \ref{fig4}. Deviation of each manipulated measurement from its steady-state value is significant, however, it is still able to avoid being discovered, due to the extremely small difference between the fed-in data and the calculated values by the SE (just around the order of $10^{-5}$ per cent) . These small values of difference greatly contribute to convincing the package for the authenticity, thus easily bypassing the BDD, even if the changes of loads at Bus 5, 6 and 8 through each case are significant. Without an additional measurement for detecting bad measurements, neither the system operator nor the program is able to reveal the existence of manipulated data. Information about all eight cases of attack is summarized in Table \ref{table1}.
\begin{table}[htbp]
	\setlength\tabcolsep{10pt}
	\caption{Results from eight different attack cases.}
	\label{table1}
	\centering
	\begin{tabular}{c|c|c|c|c}
		\toprule
		\hline
		Case & $\pm\theta_5 (^o)$ & Plausibility Check & Observability & Bad Data \\ [0.5ex] \hline
		1    & +0.5      & 0/48               & Yes           & 0/34     \\ \hline
		2    & -0.5      & 0/48               & Yes           & 0/34     \\ \hline
		3    & +1.0      & 0/48               & Yes           & 0/34     \\ \hline
		4    & -1.0      & 0/48               & Yes           & 0/34     \\ \hline
		5    & +1.5      & 0/48               & Yes           & 0/34     \\ \hline
		6    & -1.5      & 0/48               & Yes           & 0/34     \\ \hline
		7    & +2.0      & 0/48               & Yes           & 0/34     \\ \hline
		8    & -2.0      & 0/48               & Yes           & 0/34     \\ \hline
		\bottomrule	
		\end{tabular}
\end{table}

\section{FDI Data Generation processes}\label{dagen}
As demonstrated in the above experimental results, our attack design can completely bypass the existing sophisticated BDD of an SE package. From this basis, we move forward to build up the first cyber-physical AC-based FDI dataset in power system, particularly, the transmission system's dataset of state estimation (namely, TSE-DS). The object of this task is still the WECC's nine-bus system \cite{anderson2008power}. It is complicated enough to be representative, yet simple enough to build up the dataset. In addition, its loading profile is generated with the available demand data of Tasmania (an island state of Australia) that can be easily collected from Australian Energy Market Operator (AEMO) \cite{AEMO}. The TSE-DS consists of a set of normal data, which reflects the results achieved from inputting the SE with genuine measurement values, and a set of attack data, which has been collected by running the State Estimation with data containing corrupted values. The process of generating this TSE-DS is described below.

\subsection{Normal steady-state dataset}
Similar to various sample systems, the nine-bus system under investigation has only one loading data, corresponding to one steady state of operation. For this reason, a realistic demand data will be used to render different states of operation for this system. Among the available demand data from AEMO, it is the system of Tasmania State that shares a lot of similarities with the nine-bus system. Therefore, the historic demand data for Tasmania in 2018 is employed to construct various loading scenarios for the WECC nine-bus system through a process of scaling down.

After collecting enough data about load demands for the system where 52558 values of demand were created, the next task was generating genuine measurement values, which in turn will become the input data for the SE process. Although this work seems to be simple with the load flow calculation of PowerFactory, it turns out to be extremely tedious due to the large volume of input data. The scripting feature of PowerFactory, DPL (DIgSILENT Programming Language \cite{manual}) is used to automate the whole process. Corresponding to 52558 sets of load demands, there are 52558 sets of load flow results, thus the SE produced 52558 spreadsheets of SE result. For the sake of management, sheets are grouped into files with 432 sheets per file, which are equivalent to the data monitored in three consecutive days. Each spreadsheet contains the comprehensive SE results of 48 measurements in the nine-bus system. The order of measurements and the detailed information about each attribute in a spreadsheet can be found in Readme.txt file of the dataset. For every measurement, nine relevant attributes were recorded where the most important fields are the calculated measurement values and the BDD indication. In conclusion, there are 122 files containing different load demands, load flow results and state estimation results in the normal dataset. From the point of view of the intrusion detection researchers, the label for each measurement in this set is "normal".

\subsection{False Data Injection Attack dataset}
The procedure of generating attack data is partly similar to the normal data generation except it employed load flow results as ingredient to produce the attack vectors (the corrupted measurements) before injecting into the SE. The stage of attack design is described in details in the previous section with the illustrated example using only one set of steady-state values to produce eight different sets of attack measurements. Although the number of attack cases can be generated as many as we want, for every steady state corresponding to a set of load demands, two scenarios of attack are enough to create a huge volume yet complicated attack data. This dataset has similar format to the normal dataset, which contains all relevant information about every measurement in the system as well as the results from the SE. However, the SE outputs do not include explicitly any information about the attack since the proposed attack design process is able to completely bypass the BDD of the PowerFactory. There are cases that the whole set of 48 measurements were manipulated while there are cases that only part of the set were attacked. However, it can be confirmed that every spreadsheet in the folder "Attack\_TAS" contains data about attack on the SE. In conclusion, the amount of attack records is twice as much as the amount of normal data. A brief of summarization about the TSE-DS dataset is shown in Table \ref{table2}.
\begin{table}[htbp]
	\setlength\tabcolsep{15pt}
	\caption{Summary of the TSE-DS dataset.}
	\label{table2}
	\centering
	\begin{tabular}{l|l}
		\toprule
		\hline
		Description & Values \\
		\hline
		Number of records/spreadsheets                  & 157674 \\ \hline
		Number of normal records                        & 52558  \\ \hline
		Number of attack records                        & 105116 \\ \hline
		Number of attributes                            & 9      \\ \hline
		Number of measurements per sheet                & 48     \\ \hline
		Number of active power measurements per sheet   & 24     \\ \hline
		Number of reactive power measurements per sheet & 24     \\ \hline
		Number of Bad Data in normal records            & 0      \\ \hline
		Number of Bad Data in attack records            & 34-48  \\ \hline
		Number of Bad Data detected in attack records   & 0      \\ \hline
		\bottomrule
	\end{tabular}
\end{table}

\section{Preliminary analysis of the generated dataset}\label{preana}
In this part, some preliminary analyses are carried on to examine the similarity between the normal data and attack data. Since the process of State Estimation involves both physical laws of electrical system as well as the stochastic optimization, the samples of dataset (one random record in each file of normal data and its corresponding record from the attack data) are inspected under both perspectives. The inspection results demonstrate causes for the failure of the BDD in revealing injected bad measurements.
\subsection{Electrical perspective}
As stated above, the FDI attack fundamentally sets up a counterfeit steady state of operation based on collected data from a genuine state. Therefore, the deceitful state must satisfy all the physical constraints of a steady state in power system. Firstly, the power balance must be guaranteed.  It means that the total amount of active and reactive power consumed by loads plus the sum of all power losses on branches must be equal to the generated active and reactive power, respectively. In addition, status of the nodes with no power injected must be maintained. For example, an assessment based on physical system criteria for Sheet $285^{th}$ of file Normal $122^{nd}$ and its corresponding attack data is conducted. From the assessment results of the no-injection bus, the absolute maximum error for the sum of active power is around the order of $10^{-8}$ while the error for reactive power is even significantly smaller, around the order of $10^{-13}$. It means that the largest error is as large as 0.01W for such calculation. In addition, the observed results implied no difference between attack data and normal data. The second assessment results related to the mismatch between consuming and generating is just around the order of $10^{-7}$, which is only equivalent to 0.1W. Once again, the difference between normal and attack state is small enough and this looks the likely reason for the success in bypassing the BDD.

\subsection{Statistical analysis perspective}
The problem of estimating state variables is solved continuously while the bad data, if exists, will be gradually excluded from the set of input data. This process is repeated until the set of input data contains only \textit{suitable data}, not just \textit{correct data}. This is a loophole that attackers can exploit to design manipulated measurements to bypass the BDD in the SE package. More specifically, as long as the manipulated values are systematically designed to be as close as possible to the appropriate values from the point of view of the package, the BDD will consider it as the normal data. The key of this issue originates from the mechanism of the BDD.

When it comes to the measurement values, it is apparent that they contain errors due to various reasons: configuration of transducers, wiring or transmitting issues, etc. It is impossible to acquire the exact values of both measurement and the corresponding errors. However, we usually assume that the acquired values are close to the true value with only a small difference of error $\epsilon$ as below \cite{monticelli2012state}:
\begin{equation}
\label{eq15}
Z_{meas} = Z_{true} + \epsilon
\end{equation}

The values of $\epsilon$ are unknown but assumed to have a normal probability density function with zero mean. Therefore, the residual \textit{J}(\textit{x}) in Equation (\ref{eq16}) must have the probability density function as chi-square distribution \cite{wood2013power}, $\chi^2$(\textit{K}) with \textit{K} is the degree of freedom, calculated by Equation (\ref{eq17}).
\begin{equation} \label{eq16}
J(x) = \sum_{i=1}^{N_m} \frac{[Z_{meas} - Z_{calc}]^2}{\sigma_i^2}
\end{equation}
\begin{equation} \label{eq17}
K = N_m - N_s
\end{equation}
where
\begin{itemize}
	\item $\sigma_i$: Standard deviation of measurement $i^{th}$.
	\item $N_m$: Number of measurements in the system.
	\item $N_s$: Number of states, equal to $(2n - 1)$ with \textit{n} is the number of nodes in the system.
\end{itemize}

The mechanism of BDD is simply by comparing the value of residual \textit{J}(\textit{x}) with the value of the chi-square distribution for certain value of \textit{K} at a significance level. The chosen value of significance level implies the false alarm occurrence created by the process of hypothesis testing \cite{wood2013power}. Likewise, a significance value of 0.005 means only 0.5 per cent cases raise a false alarm. DIgSILENT does not reveal the exact value of significance level in their SE package. Thus, in order to reduce the false alarm rate in a large data volume environment, in companion with the parameters of our current work $N_m$ = 48 and $N_s$ = 17, a significance level of 0.005 was chosen, resulting in the threshold to identify bad data being 53.67. It means that whichever set of input data that can bring out the residual of greater than 53.67 will be suspected to contain bad data and vice versa. The computation of \textit{J}(\textit{x}) for generated data was conducted on the same samples from the electrical perspective. For instance, with data from Sheet $285^{th}$ of file Normal $122^{nd}$ and its corresponding attack data, both \textit{J}(\textit{x}) values for attack ($\approx0.00224$) and normal data ($\approx2.21\times10^{-8}$) are small enough to avoid being marked as suspected, which is far below the threshold of 53.67. The same calculation procedure is conducted for one random record in each files of normal data and its corresponding record from the attack data, yielding the value of residual fluctuated around the order of $10^{-3}$ to $10^{-8}$. These miniscule values making the set of attack data having the full reputation as being the normal data. Therefore, the failure of the BDD is inevitable due to the similarities in both physical and statistical aspects of two types of dataset.

\section{Conclusion and Future work}\label{conclu}
In this paper, an AC model based FDI attack design scheme is proposed. A case study with industrial standard AC-based SE is used to demonstrate the scheme. The BDD, which is specifically equipped for the purpose of detecting and eliminating any inappropriate data, completely failed to notice the existence of manipulated measurements in the set of input data, even when they make up a majority of the data. This work distinguished itself from various works about DC-based SE, which is only a simplified linearized version of the AC counterpart. Several works in Section \ref{prevwork} also brought out the poor performance of DC-based SE systems. We have shown that the DC-based FDI attack model where most of existing works are based on is unlikely successful in passing the AC-based commercial-scale BDD mechanism.

Traditionally, State Estimation is deployed for the transmission system only. However, as the integration of Distributed Energy Resources quickly spreads out in recent years, the implementation of SE in the distribution system is indispensable. The newly emerged distribution system will then have a much more sophisticated structure with bi-directional power flows. Subsequently, the system monitoring work will be more complicated than in the classical distribution system with radial feeder only. Thus, the security considerations in distribution systems are indispensable for future work. Since the structure of distribution system is getting closer and closer to the transmission system (meshed, various loads and sources, etc.), there exists an expectation that our proposed attack design is applicable to active distribution as well as undetectable with the existing commercial BDD mechanisms. Therefore, our cyber-physical dataset will be a valuable resource for the purpose of conducting research on system's defense strategies. The FDI dataset TSE-DS can be acquired by sending email to the corresponding author Prof. Jiankun Hu at J.Hu@adfa.edu.au.

\section*{Acknowledgements}
This research is supported by ARC Discovery Grant (IDDP190103660) 561 and ARC Linkage Grant (IDLP180100663).

\bibliography{B1,wileyNJD-AMA}%

\end{document}